\magnification\magstep1
\baselineskip=12pt
\parindent=3truepc
\def\bigskip{\vskip 6 truemm}
\def\medskip{\vskip 4truemm}

\input epsf.tex
\font\rfont=cmr10 at 10 true pt
\def\ref#1{$^{\hbox{\rfont {#1}}}$}


\font\mathten=cmsy10 scaled\magstep0

\font\tenit=cmti10 scaled\magstep0

   
\def\e{\epsilon}

\def\pmb#1{\setbox0=\hbox{#1}
 \kern.05em\copy0\kern-\wd0 \kern-.025em\raise.0433em\box0 }


 %


\def\boxit#1{\vbox{\hrule\hbox{\vrule\kern1pt\vbox
{\kern1pt#1\kern1pt}\kern1pt\vrule}\hrule}}

\def\h{\hfill\break}
\parskip=6pt
\parindent=0pt
\hsize=6truein\hoffset=-5truemm
\voffset=-1truecm\vsize=8.5truein
\def\footnoterule{\kern-3pt
\hrule width 17truecm \kern 2.6pt}
\def\h{\hfill\break}


\catcode`\@=11 

\def\nolabels{\def\wrlabeL##1{}\def\eqlabeL##1{}\def\reflabeL##1{}}
\def\writelabels{\def\wrlabeL##1{\leavevmode\vadjust{\rlap{\smash%
{\line{{\escapechar=` \hfill\rlap{\sevenrm\hskip.03in\string##1}}}}}}}%
\def\eqlabeL##1{{\escapechar-1\rlap{\sevenrm\hskip.05in\string##1}}}%
\def\reflabeL##1{\noexpand\llap{\noexpand\sevenrm\string\string\string##1}}}
\nolabels
\global\newcount\refno \global\refno=1
\newwrite\rfile
\def\defref{$^{{\hbox{\rfont \the\refno}}}$\nref}
\def\nref#1{\xdef#1{\the\refno}\writedef{#1\leftbracket#1}%
\ifnum\refno=1\immediate\openout\rfile=refs.tmp\fi
\global\advance\refno by1\chardef\wfile=\rfile\immediate
\write\rfile{\noexpand\item{#1\ }\reflabeL{#1\hskip.31in}\pctsign}\findarg}
\def\findarg#1#{\begingroup\obeylines\newlinechar=`\^^M\pass@rg}
{\obeylines\gdef\pass@rg#1{\writ@line\relax #1^^M\hbox{}^^M}%
\gdef\writ@line#1^^M{\expandafter\toks0\expandafter{\striprel@x #1}%
\edef\next{\the\toks0}\ifx\next\em@rk\let\next=\endgroup\else\ifx\next\empty%
\else\immediate\write\wfile{\the\toks0}\fi\let\next=\writ@line\fi\next\relax}}
\def\striprel@x#1{} \def\em@rk{\hbox{}} 
\def\lref{\begingroup\obeylines\lr@f}
\def\lr@f#1#2{\gdef#1{\defref#1{#2}}\endgroup\unskip}
\newwrite\lfile
{\escapechar-1\xdef\pctsign{\string\%}\xdef\leftbracket{\string\{}
\xdef\rightbracket{\string\}}}

\def\writestop{\def\writestoppt{\immediate\write\lfile{\string\p
ageno%
\the\pageno\string\startrefs\leftbracket\the\refno\rightbracket%
\string\def\string\secsym\leftbracket\secsym\rightbracket%
\string\secno\the\secno\string\meqno\the\meqno}\immediate\closeout\lfile}}
\def\writestoppt{}\def\writedef#1{}
\catcode`\@=12 
\rightline{DAMTP 96/47}
\vskip 5truemm
\centerline{THE BFKL POMERON: CAN IT BE DETECTED?}
\vskip 10truemm
\centerline{P V LANDSHOFF}
\centerline{\tenit DAMTP, University of Cambridge}
\centerline{\tenit Cambridge CB3 9EW, England}
\vskip 10truemm
\centerline{ABSTRACT}
\vskip -2mm

\midinsert\leftskip 3truepc\rightskip 3truepc{{\rfont\noindent
The BFKL pomeron is swamped by the soft pomeron, at least at ${\mathten t=0}$.
}}\endinsert
\bigskip\rm

Figure 1 shows data for the $pp$ and $\bar pp$ total cross-sections. The 
curves\defref\ad{
A Donnachie and P V Landshoff, Physics Letters B296 (1992) 227
}
corrrespond to the exchange of the $\rho ,\omega, f, a$ trajectories, whose
contribution falls with increasing energy approximately as $1/\surd s$,
and a rising soft-pomeron-exchange term which rises as $s^{0.08}$.
There is a clear disagreement between the two Tevatron data points at
$\surd s=1800$ GeV. If one believes the higher CDF measurement\defref\cdf{
CDF collaboration: F Abe et al, Physical Review D50 (1994) 5550
}, rather than the lower E710 one\defref\e710{
E710 collaboration: N Amos et al, Phys Lett B243 (1990) 158
}, then there is room for an additional contribution of at most 10 mb at
that energy. This is the  limit on how large any additional contribution 
can be at that energy.

One such contribution might be from a second pomeron. In particular, the
BFKL pomeron\defref\bfkl{
E A Kuraev, L N Lipatov and V Fadin, Soviet Physics JETP 45 (1977) 199\h
Y Y Balitskii and L N Lipatov, Sov J Nuclear Physics 28 (1978) 822
} is thought to give a contribution that rises as fast as $s^{0.3}$ or
more. The BFKL pomeron is purely perturbative, and so it is often said
that it is not applicable to purely soft processes such as hadronic
total cross-sections.  However, a more correct statement is that in soft 
processes perturbative contributions are
swamped by nonperturbative ones. Nevertheless they are
present, and the data in figure 1 limit how large they can be. This in
turn\defref\cudell{
J R Cudell, A Donnachie and P V Landshoff, hep-ph/9602284
} limits how large they can be in the hard processes where they might be
expected to dominate.

Analyses of the BFKL equation often incorrectly 
extend the integrations over the 
loop momenta to all values. If this is done, the separate terms in the
BFKL equation are infrared divergent, but the divergences cancel between
the terms. Nevertheless it is illegal to allow the integration to extend into
the infrared region, because this is the nonperturbative region and the 
BFKL equation is purely perturbative. Likewise, it is not legal to allow
arbitrarily large loop momenta, because this violates energy conservation.

\topinsert
\centerline{{\epsfxsize=130truemm\epsfbox{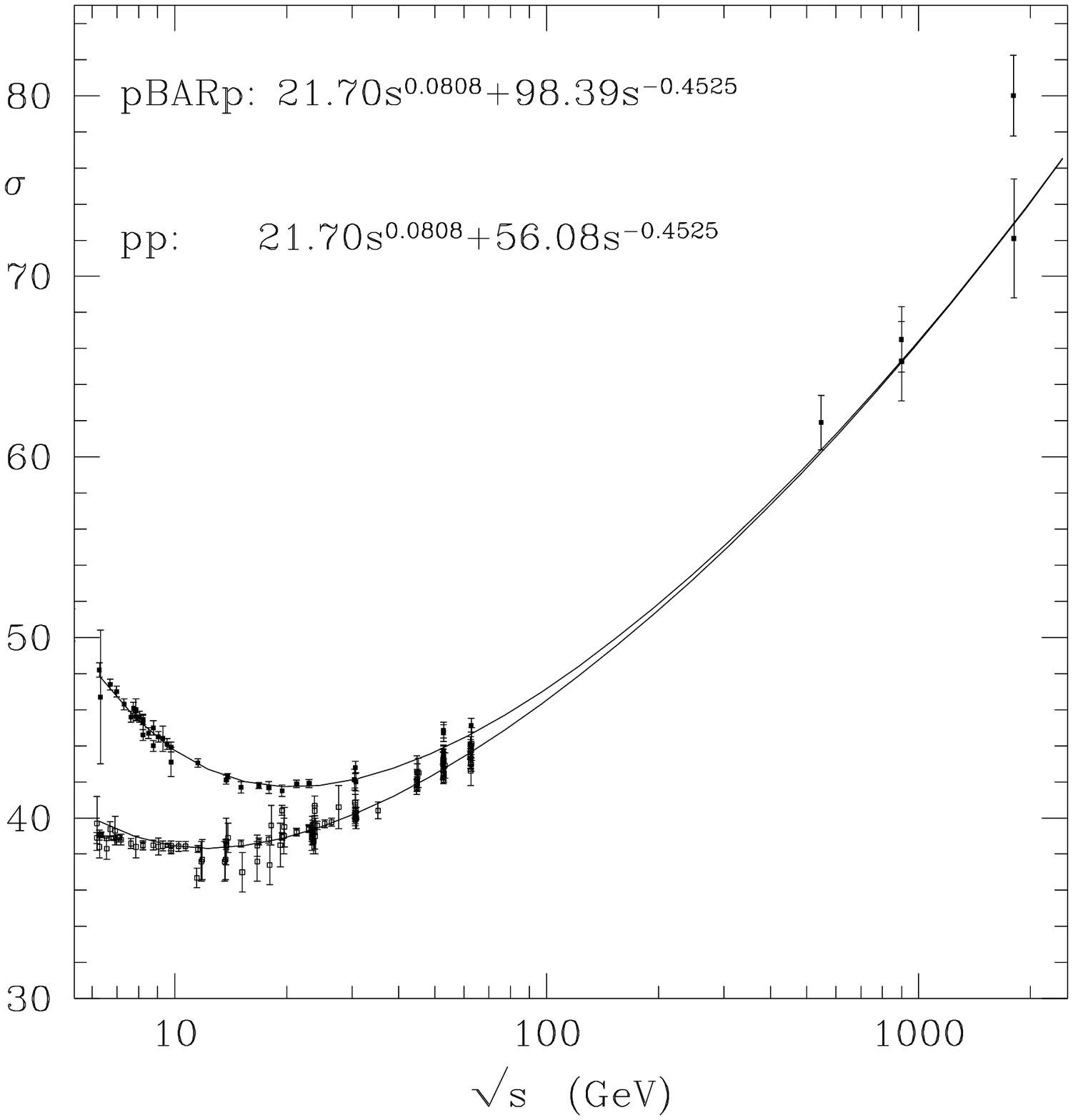}}}\hfill
\vskip -5mm
\centerline{\rfont Figure 1: Data for $pp$ and $\bar pp$ total cross-sections, 
with the fits indicated}
\endinsert

\midinsert
\centerline{{\epsfxsize=80truemm\epsfbox{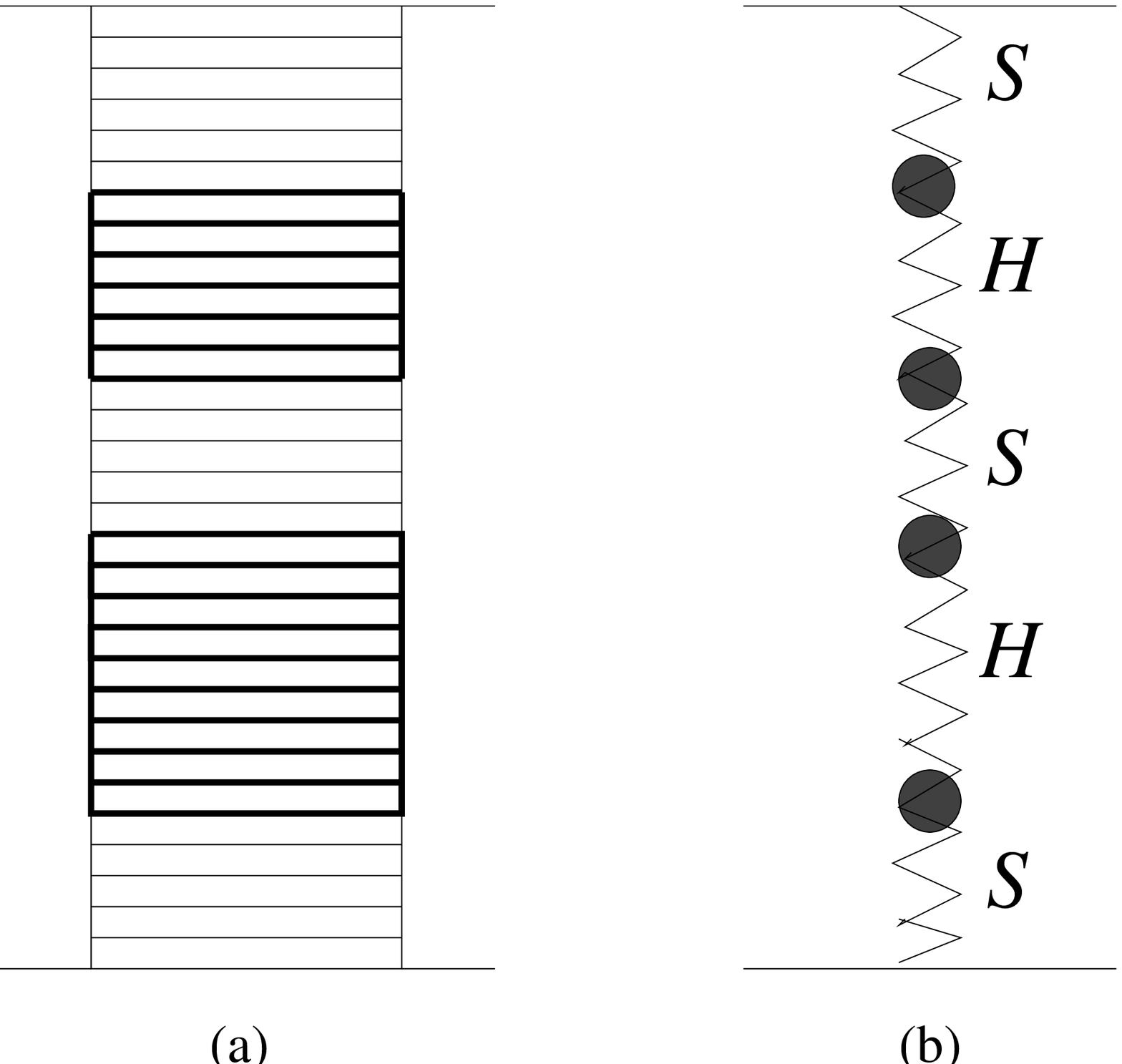}}}\hfill
\vskip -5mm
{\rfont Figure 2: (a) alternating groups of partons with low and high $K_T$,
with (b) their sum giving alternating soft and hard pomerons.}
\bigskip
\bigskip
\centerline{\rfont{\epsfxsize=60truemm\epsfbox{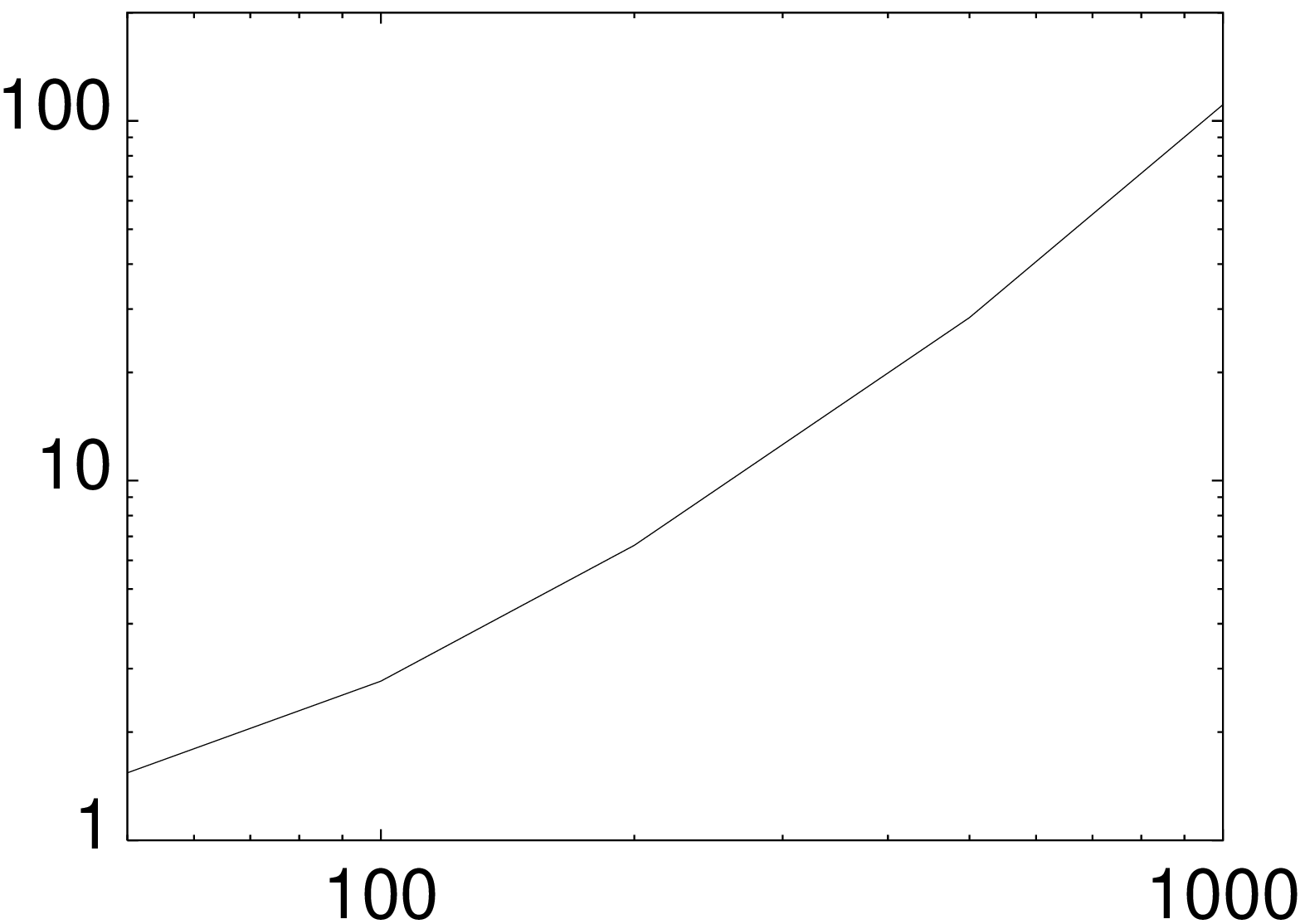}}}\hfill\break
\vskip -10mm
\centerline{${\mathten\surd s}$}
\vskip 5mm
\centerline{\rfont Figure 3: $\sigma _{qq}(K_T>\mu)$ in microbarns for $\mu =2$ GeV}
\endinsert

The BFKL equation describes the emission of partons. To avoid the
nonperturbative problems, we have to place some lower limit $\mu$
on their transverse momentum if we are to believe its predictions. 
In a general event, we may group  the final-state partons according to
their rapidities. As there is no transverse-momentum ordering, their
transverse momentum is not correlated with their rapidity. So as we
scan the rapidity range we find groups of partons all having transverse
momentum greater than $\mu$, with each such group separated by a group
in which none of the partons has transverse momentum greater than ${\mu}$.
This we show in figure 2a, where the heavy lines have transverse momentum
${K_T>\mu}$, while the light lines have ${K_T<\mu}$.
When we sum over all possible numbers of lines in a group with
${K_T>\mu}$ we obtain the hard pomeron $H$ which we may calculate
from the BFKL equation, while a group with ${K_T<\mu}$ sums to a soft exchange
$S$. So the result is figure 2b.
When we sum over all final states, 
we obtain for the cross-section
$$
S+H+SH+HS+SHS+\dots
\eqno(1)
$$
Obviously the sum must be independent of the value we have chosen for
$\mu$, provided only that $\mu$ is large enough for the perturbative
BFKL equation to be applicable to the calculation of $H$. It turns 
out\ref{\cudell} that $\mu$ must be at least 2 GeV in order that the hard
contribution to the $\bar pp$ cross-section shall not exceed the 10 mb
limit at Tevatron energy. 

This is shown in figure 3, which is the calculated BFKL contribution $H$
to the quark-antiquark total cross-section for the choice $\mu =2$ GeV.
It must be multiplied by 9 to get the contribution to the 
$\bar pp$ cross-section. Adding in the terms $SH+HS+SHS+\dots$ mutliplies
it by a factor which we have estimated to be at most an order of magnitude.
At the quark level, the 1800 GeV Tevatron energy corresponds to 600 GeV,
and the value of $\sigma _{qq}(K_T>\mu)$ at this energy 
shown in figure 3 is about
as large as can be without conflicting with the data in figure 1.
Notice that, if we had not required the total transverse energy of the emitted
partons to be less than $\surd s$, the output for $\sigma _{qq}(K_T>\mu)$
would have been an order of magnitude larger.

Figure 4 shows the lowest-order contributions to the process
$\gamma ^* q\to\rho q$. As $Q^2$ increases, the two diagrams cancel
each other more and more, a property known as colour transparency.
The result of making extra perturbative insertions in the diagrams
through the BFKL equation is shown in figure 5, again for the choice
$\mu =2$ GeV. This figure shows also the soft-pomeron-exchange 
contribution\defref\rrho{
A Donnachie and P V Landshoff, Physics Letters B185 (1987) 403
}, which fits well to fixed-target data\defref\nmc{
NMC Collaboration, P Amaudruz et al, Z. Physik C54 (1992) 239
} and the H1 data from HERA (though ZEUS finds\defref\zeus{
ZEUS Collaboration, M Derrick et al: DESY 95-133
} a slightly larger cross-section). As may be seen, the BFKL contribution 
is some 2 orders of magnitude too small {\it in the amplitude} to
explain the data.

\topinsert
\centerline{\rfont{\epsfxsize=120truemm\epsfbox{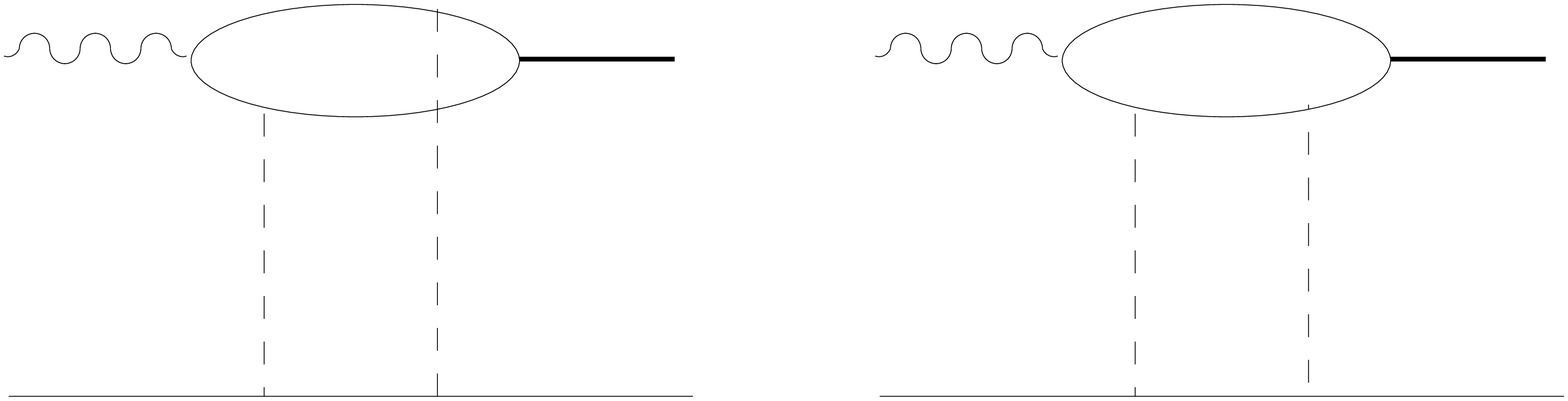}}}\hfill\break
\centerline{\rfont Figure 4: Lowest-order graphs for {\mathten
 $\gamma ^*q\to\rho q$}}
\vskip 10mm
\centerline{\rfont{\epsfxsize=60truemm\epsfbox{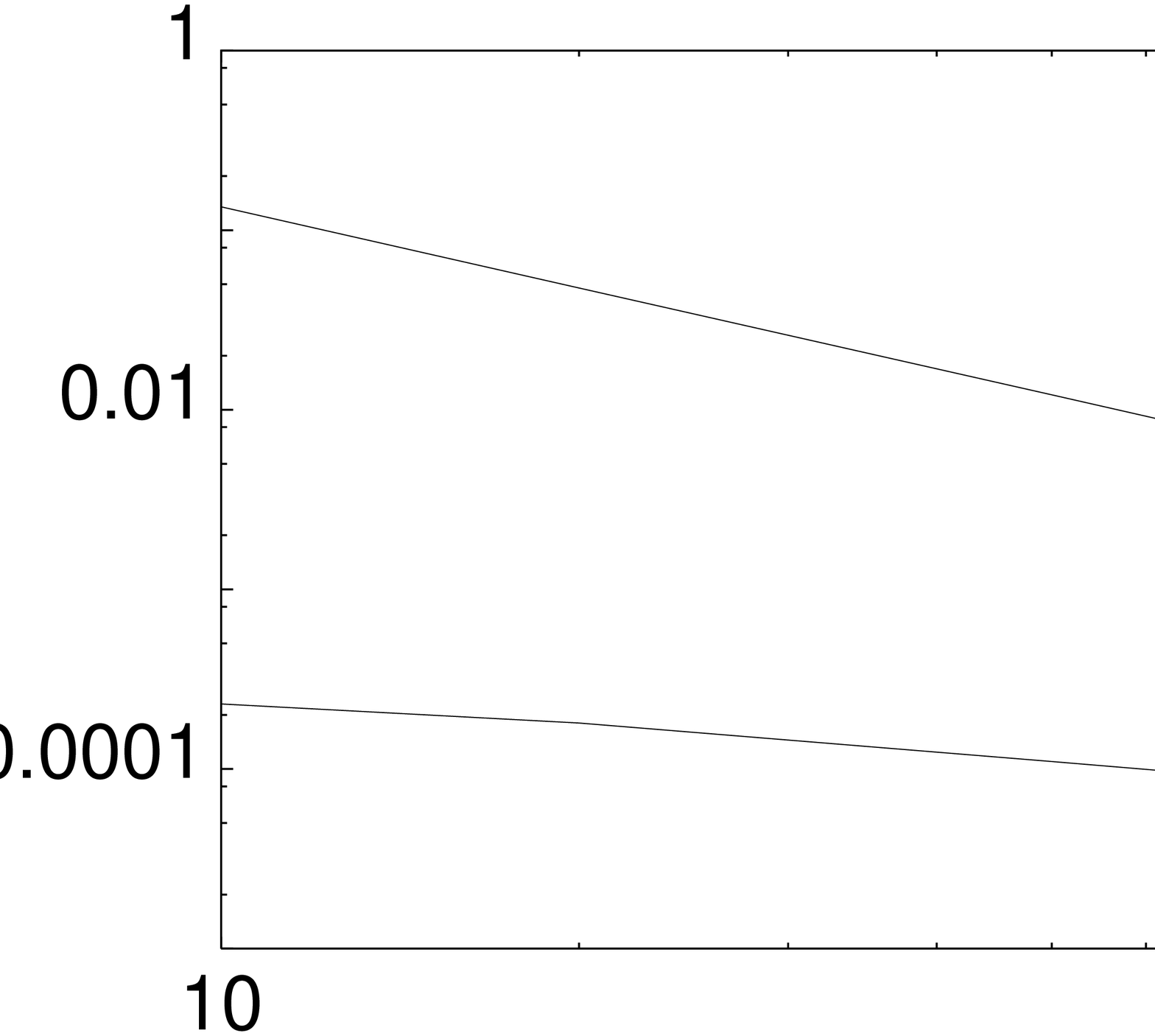}}}\hfill
\vskip -5truemm
\centerline{$\surd s$}
{\rfont Figure 5: Pomeron-exchange contributions to the amplitude
for ${\mathten \gamma ^* q\to\rho q}$; the upper curve corresponds 
to the soft pomeron
and the lower to the BFKL pomeron}
\endinsert

Although the BFKL contribution is so very small, its properties
are more or less as expected. For example, for $\surd s=100$ GeV,
reducing $\mu$ from 2 GeV to 1 GeV causes a huge increase in
the amplitude for the soft process $qq\to qq$ --- some two orders of
magnitude. At the same energy and at $Q^2=1000$ GeV$^2$ the increase
is only a factor of 5. This property is called diffusion\defref\bartels{
J Bartels, H Lotter and M Vogt, Physics Letters B373 (1996) 215
}: the effects of the hard interaction at the top of the BFKL ladder
are felt all the way down it. For the amplitude $\gamma ^*\gamma ^*\to\rho\rho$,
where there is a hard interaction at both ends of the ladder,
the effect is even more marked\ref{\cudell}: the factor increase reduces to 3.

\bigskip\bigskip
{\it This research is supported in part by the EU Programme ``Human Capital
and Mobility", Network ``Physics at High Energy Colliders'', contract
CHRX-CT93-0357 (DG 12 COMA), and by PPARC.}
\bigskip
\bigskip\medskip\immediate\closeout\rfile\writestoppt
\baselineskip=10pt{{\bf References}}\bigskip{\frenchspacing%
\parindent=3truepc\escapechar=` \input refs.tmp\bigskip}\nonfrenchspacing
\bye